\PassOptionsToPackage{table}{xcolor}
\documentclass[conference]{IEEEtran}
\IEEEoverridecommandlockouts

\def\BibTeX{{\rm B\kern-.05em{\sc i\kern-.025em b}\kern-.08em
		T\kern-.1667em\lower.7ex\hbox{E}\kern-.125emX}}

\usepackage{nicefrac}
\usepackage[switch]{lineno}
\usepackage{comment}

\usepackage{float}
\usepackage{fancybox,framed}

\usepackage{booktabs} 
\usepackage{xspace}
\usepackage{tabularx,colortbl}
\usepackage{multirow, makecell}
\usepackage{paralist}
\usepackage{tabularx}
\usepackage{listings}
\usepackage{graphicx}
\usepackage{ifthen}
\usepackage{indentfirst}
\usepackage{xcolor}

\usepackage{hyperref}

\definecolor{darkjunglegreen}{rgb}{0.000000,0.392157,0.000000}

\usepackage{csquotes}
\newcommand{\nametool}{\textsc{AdaptDroid}\xspace}
\newcommand{\toolname}{\nametool}
\newcommand{\tool}{\nametool}

\definecolor{darkjunglegreen}{rgb}{0.000000,0.392157,0.000000}

\newcommand{\hide}[1]{}
\usepackage{url}
\def\codesize{\small}
\def\<#1>{\codeid{#1}}\protected\def\codeid#1{\ifmmode{\mbox{\codesize\ttfamily
			#1}}\else{\codesize\ttfamily
		#1}\fi}

\usepackage{url}
\def\codesize{\small}
\def\<#1>{\codeid{#1}}\protected\def\codeid#1{\ifmmode{\mbox{\codesize\ttfamily
			#1}}\else{\codesize\ttfamily
		#1}\fi}

\definecolor{mygray}{gray}{0.9}
\usepackage{fancybox}
\usepackage{listings}
\usepackage{etoolbox}

\makeatletter
\newcommand{\removelatexerror}{\let\@latex@error\@gobble}
 
\let\textquotedbl="

\usepackage{booktabs} 
\usepackage{graphics} 
\usepackage{multirow,bigstrut}
\usepackage[framemethod=tikz]{mdframed}
\usepackage{graphicx}
\usepackage{relsize}

\newcommand*{\all}{et al. }
\usepackage{multirow,array,varwidth}
\usepackage{booktabs}

\usepackage{listings}
\usepackage{color,url}

\definecolor{javared}{rgb}{0.6,0,0} 
\definecolor{javagreen}{rgb}{0.25,0.5,0.35} 
\definecolor{javapurple}{rgb}{0.5,0,0.35} 
\definecolor{javadocblue}{rgb}{0.25,0.35,0.75} 
\usepackage{lipsum}
\usepackage{paralist}

\usepackage{amsmath,amssymb,latexsym}

\definecolor{orange}{rgb}{1,0.5,0}
\definecolor{darkjunglegreen}{rgb}{0.000000,0.392157,0.000000}

\newcommand{\quotes}{\textquotesingle\textquotesingle}

\newcommand{\F}{{\small \textsf{false}}\@\xspace}

\newcommand{\T}{{\small \textsf{true}}\@\xspace}

\usepackage{graphicx}
\newsavebox\CBox
\def\textBF#1{\sbox\CBox{#1}\resizebox{\wd\CBox}{\ht\CBox}{\textbf{#1}}}

\usepackage{paralist}

\usepackage[shortlabels]{enumitem}
\usepackage{amsmath}
\usepackage{cite}

\begin{document}
	
\title{An Evolutionary Approach to \\ Adapt Tests Across Mobile Apps}

\author{
	\IEEEauthorblockN{Leonardo Mariani\IEEEauthorrefmark{1}, Mauro Pezzè\IEEEauthorrefmark{2}\IEEEauthorrefmark{3}, Valerio Terragni\IEEEauthorrefmark{2} and Daniele Zuddas\IEEEauthorrefmark{2}}
	\IEEEauthorblockA{\IEEEauthorrefmark{1}University of Milano Bicocca, Milan, Italy}
	\IEEEauthorblockA{\IEEEauthorrefmark{2}Università della Svizzera italiana, Lugano, Switzerland}
	\IEEEauthorblockA{\IEEEauthorrefmark{3}Schaffhausen Institute of Technology, Schaffhausen, Switzerland \\
		 leonardo.mariani@unimib.it - \{mauro.pezze, valerio.terragni, daniele.zuddas\}@usi.ch
	 }\\ 
\vspace{-4mm}
\emph{This is the author's version of the work. The definitive version appeared at AST 2021} 

}

\maketitle

\thispagestyle{plain}
\pagestyle{plain}

\begin{abstract}
Automatic generators of GUI tests often fail to generate semantically relevant test cases, and thus miss important test scenarios.
To address this issue, test adaptation techniques can be used to automatically generate semantically meaningful GUI tests from test cases of applications with similar functionalities.

In this paper, we present \tool, a technique that approaches the test adaptation problem as a search-problem, and uses evolutionary testing to adapt GUI tests (including oracles) across similar Android apps.
In our evaluation with 32 popular Android apps, \tool successfully adapted semantically relevant test cases in 11 out of 20 cross-app adaptation scenarios.
\end{abstract}


\vspace{3mm}
\section{Introduction}

Verifying GUI applications is both important, due to their pervasiveness, and challenging, due to the huge size of their execution space~\cite{Mirzaei:combinatorialGUI:ICSE:2016}. 
GUI testing is a popular way to verify the behavior of GUI applications, which  amounts to design and execute GUI test cases.
A GUI test case (GUI test in short) consists of 
\begin{inparaenum}[(i)]
	\item a sequence of events that interact with the GUI, and 
	\item assertion oracles that predicate on the GUI state.
\end{inparaenum}

Because manually designing GUI tests is expensive, many 
automatic GUI test generators have been proposed.
Current approaches generate GUI tests either randomly~\cite{machiry:dynodroid:FSE:2013} or by relying on structural information that they obtain either from the GUI~\cite{Memon:GUIRipping:WCRE:2003, Mirzaei:sigdroid:ISSRE:2015,su:modelbasedgui:FSE:2017} or from the source code~\cite{Amalfitano:AndroidRipping:ICSE:2012}. 
Current approaches suffer from two main limitations. 
By largely ignoring the semantics of the application, they produce  tests that are either semantically meaningless or unrepresentative of the canonical usage of the application~\cite{Yang:gui:JSS:2014}.
Thus, they likely miss the GUI event sequences that properly exercise functionalities and reveal faults~\cite{bozkurt:realistic:sose:2011,choudhary:android:ASE:2015}.
Moreover, current GUI test generators rely on implicit oracles~\cite{Moran:crashes:ICST:2016,Zhao:crash:ICSE:2019} that miss many failures related to semantic issues~\cite{zeng:automated:fse:2016,choudhary:android:ASE:2015}.

Recently, researchers investigated the opportunity to address these challenges by
exploiting semantic similarities across GUI applications~\cite{rau:efficent:icse:2018, Mariani:Augusto:ICSE:2018,Hu:appflow:FSE:2018}.
Indeed, browsing  the Google Play Store  reveals many Android apps that are semantically similar, albeit offering different graphics appearance, access permissions, side features, and user experience~\cite{Linares:similarapps:icpc:2016,Ruiz:reuse:ieee:2014,Li:similarities:JCST:2019}.
Hu \all have shown that among the top 309 non-game mobile apps in the Google Play Store, 196 (63.4\%) of them fall into 15 groups that share many common functionalities~\cite{Hu:appflow:FSE:2018}.
This confirms the huge potential of sharing tests across similar applications because common functionalities yield to common GUI tests~\cite{Hu:appflow:FSE:2018}.

The recent \textsc{CraftDroid}~\cite{lin:craftdroid:ASE:2019}  and \textsc{AppTestMigrator}~\cite{behrang:apptestmigrator:ASE:2019} approaches generate GUI tests
by automatically adapting existing GUI tests across similar Android apps. 
Both approaches generate new tests for a \textit{recipient app}, by adapting the tests designed for some \emph{donor app} that shares semantically similar functionalities with the \textit{recipient app}.
Test adaptation, when successful, addresses the limitations of existing GUI test generators:
\begin{inparaenum}[(i)]
\item it yields to semantically meaningful GUI tests that characterize canonical usages of the app under test.
\item  it leverages the functional oracles of the donor tests.
\end{inparaenum}

\textsc{CraftDroid} and \textsc{AppTestMigrator} explore a GUI model of the recipient app to find a sequence of events that maximize the semantic similarity with the events of the donor test.
They compute the semantic similarity of GUI events using word embedding~\cite{Mikolov:VectorSum:NIPS:2013} applied to the  textual descriptors of events extracted from the GUI widgets.
Both techniques \emph{greedily} explore a single test adaptation scenario, missing the many alternative adapted tests that could be generated starting from a same donor test. 
%
%
Indeed, extensively exploring the execution space is often imperative to identify a sequence of events that well reflects the semantics of the donor test.

In this paper, we present \tool, a technique that formulates the GUI test adaptation problem as a search-problem using an evolutionary approach.
\tool explores the huge space of GUI tests with a fitness function that rewards the tests that are most similar to the donor test.
The \tool notion of similarity 
considers both the semantics of the events and the capability of the adapted test to reach states where the donor oracle can be applied to. 

We implemented \tool in a prototype tool, and evaluated with a human study involving  32 Android apps. 
Our results show that \tool successfully adapts semantically relevant GUI tests in 11 out of 20 test adaptation scenarios. Thus confirming that test adaptation is a promising and complementary solution for generating GUI tests.

\begin{figure*}[th]
	
	\centering
	\resizebox{1\textwidth}{!}{%
		\includegraphics{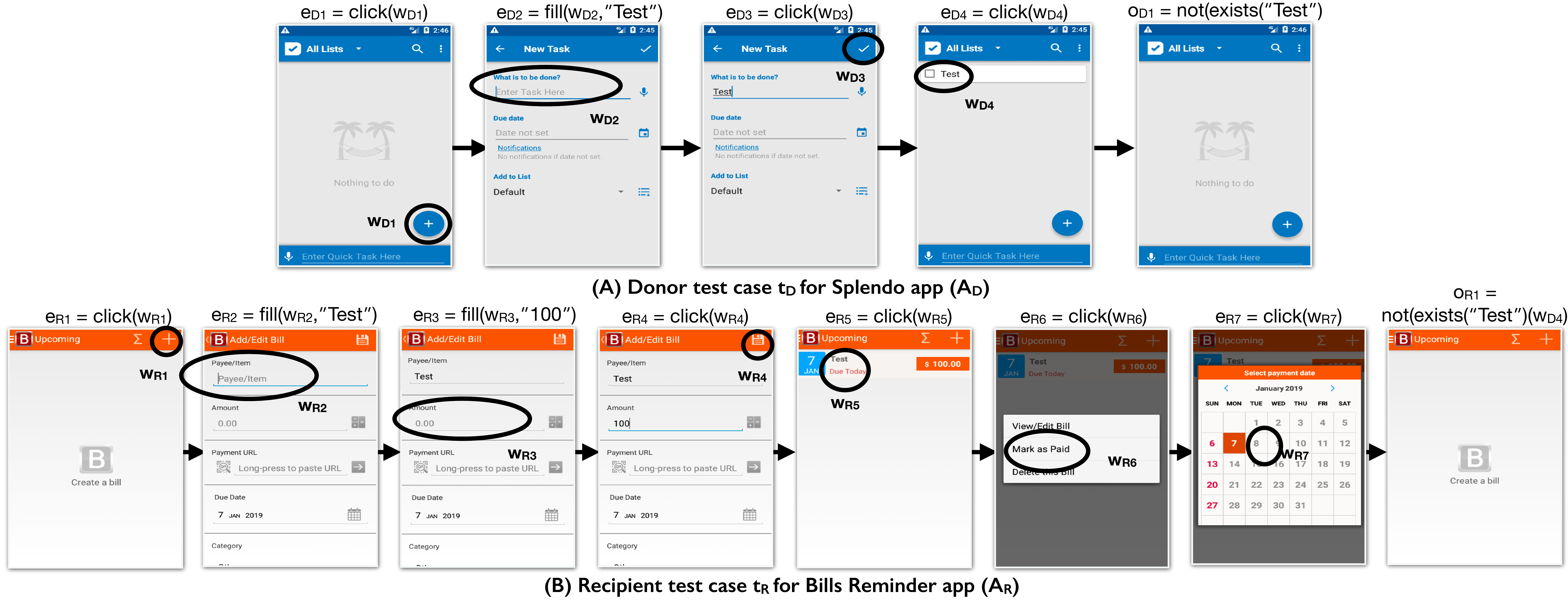}
	}
		\vspace{-6mm}
\caption{Example of \nametool cross-application test adaptation}
\label{adaptExample}
\vspace{-3mm}

\end{figure*}

In summary, the main contributions of this paper are:

\begin{itemize}[leftmargin=*]
	\item formulating the problem of adapting GUI tests across similar applications as an evolutionary approach,
	\item proposing \nametool, to adapt both GUI event sequences and oracle assertions across mobile apps,
	\item presenting the results of a study showing that test adaptation of Android apps is a valuable opportunity,
	\item presenting an empirical evaluation of \tool  that highlights its effectiveness and limitations,
	\item releasing the \nametool tool and all experimental data~\cite{adaptdroid}.
\end{itemize}


\section{Adapting Tests Across GUI Applications}


\label{sec:preliminaries}

GUI applications  interact with users through a Graphical User Interface (GUI)~\cite{dix:HCI:2009}.
A \textBF{GUI} is a forest of hierarchical windows (\emph{activities} in Android),
where only one window is active at any time~\cite{Memon:GUIRipping:WCRE:2003}. 
Windows host \textBF{widgets}, which are atomic GUI elements characterised by
properties: \emph{type}, displayed \emph{text}  (if any) and \emph{xpath} (a label that uniquely identifies the widget in the structural hierarchy of the window~\cite{Song:xpath:wisa:2017}).
At any time, the active window has a \textBF{state} $\mathbf{S}$ that encompasses the state (properties values) of the displayed
widgets. 
Some widgets expose user-actionable events that users can trigger to interact with the GUI.
For instance, users can click on widgets of type button or can fill widgets of type text field. 

\smallskip
A \textBF{GUI test} $\mathbf{t}$ is an ordered sequence of events  $ \langle e_1,..., e_n\rangle$ on widgets of the active windows. A test execution induces a sequence of \emph{observable} state transitions \( S_{0} \xrightarrow{e_1} S_1 \xrightarrow{e_2} S_2 \ldots
\xrightarrow{e_n} S_{n} \), where $S_{i-1}$ and $S_i$ denote the states of the active window before and after the execution of event $e_i$, respectively.  
An event is an atomic interaction on a widget. Events are typed. In this paper, we consider two common types of events
\begin{itemize}[-,leftmargin=*]
	\item \mbox{\textit{\textbf{click}($w$)}}: clicking a widget $w$;
	\item \mbox{\textit{\textbf{fill}($w$, txt)}}: filling a string \emph{txt} in widget $w$.
\end{itemize}

Each test $t$ is associated with one or more \textbf{assertion oracles}~\cite{barr:survey:tse:2015} that check the correctness of the
state $S_n$ obtained after the execution of $t$~\cite{Zaeem:oracle:ICST:2014}.
We use $O_t$ to denote the assertions associated with the test $t$, and consider two types of assertions: 
 \begin{itemize}[-,leftmargin=*]
\setlength\itemsep{0em}
	\item \emph{\textbf{exists}(txt)} checks if $S_n$  contains a widget with text \emph{txt}: \textit{exists(txt)} $ \Leftrightarrow \exists w \in S_n : \textit{text(w)}= $ \textit{txt};

\item  \textit{\textbf{hasText}(w, txt)} checks if $S_n$ has a widget $w$ with text \textit{txt}: \textit{hasText(w, txt)} $ \Leftrightarrow \exists w \textquoteright \in S_n : \textit{xpath(}w \textquoteright \textit{)} = \textit{xpath(}w\textit{)}$  $\land $ $\textit{text(}w \textquoteright  \textit{)} = $ \textit{txt}.
	
\end{itemize}

This paper presents \tool to adapt GUI tests (including oracles) across interactive applications that implement similar functionalities. Given two Android apps $A_{D}$ (donor), $A_{R}$ (recipient), and a ``donor'' test $t_{D}$ for $A_{D}$, \nametool generates a ``recipient''  test $t_{R}$ that tests $A_R$ as $t_D$ tests $A_{D}$.

\section{Working Example}
\label{example}

Figure~\ref{adaptExample} introduces a working example that illustrates the challenges of adapting GUI tests across similar applications.
Figure~\ref{adaptExample}A shows a donor GUI test ($t_D$) designed for \href{https://play.google.com/store/apps/details?id=com.splendapps.splendo}{\emph{Splendo}}, an Android app to manage tasks lists.
The test  \emph{adds a new task to a task list, and verifies that the task disappears once marked as done}. 
Figure~\ref{adaptExample}B shows how \toolname successfully adapts $t_D$ to the recipient app \href{https://play.google.com/store/apps/details?id=com.amazier.apps.billsreminder}{\emph{Bills Reminder}} ($A_R$), by generating $t_R$ that \emph{adds a new bill to the bill list and verifies that the bill disappears once marked as paid}. 
Although the two apps belong to different domains, they share
the logical operations of creating a new element (a task in $A_D$, a bill in $A_R$) and marking it as completed (done in $A_D$, paid in $A_R$).
Automatically adapting GUI tests across apps presents three main challenges: 


\smallskip
\noindent
\mbox{\textBF{1) Huge space of GUI tests}} The space of the possible GUI tests grows exponentially
with both the length of the donor test and the number of widgets in the recipient app~\cite{Mirzaei:combinatorialGUI:ICSE:2016}.  
Adapting tests requires an effective search strategy that  recognizes the relevant GUI events in the recipient app. 
	
\smallskip
\noindent
\mbox{\textBF{2) GUI differences}} The donor test may exercise GUI widgets that are logically
equivalent but very different from the widgets of the recipient app. 
For instance in Figure~\ref{adaptExample}, semantically similar widgets are labelled  \quotes\texttt{What is to be done?}\quotes $ $ ($w_{D2}$) and
\quotes\texttt{Payee/Item}\quotes $ $  ($w_{R2}$), respectively.  Also, \emph{Splendo} uses a tick mark button ($w_{D3}$) to save a task, while \emph{Bills Reminder} uses a floppy disk image button ($w_{R4}$).

	
\smallskip
\noindent
\mbox{\textBF{3) No one-to-one GUI event matching}} The donor and adapted tests might have a different number of events. 
For instance, in Figure~\ref{adaptExample} the donor and recipient tests have four and seven events, respectively.
Creating a bill in \emph{Bills~Reminder} requires more events than
creating a task in \emph{Splendo}. 
Marking a bill as paid in \emph{Bills~Reminder} requires a date, while marking a task as done in \emph{Splendo} does not. 



\smallskip
The next Section presents \nametool and discusses how it addresses these challenges.

\section{\nametool}

\begin{figure}[t!]

	\centering
	\resizebox{1\linewidth}{!}{%
		\includegraphics{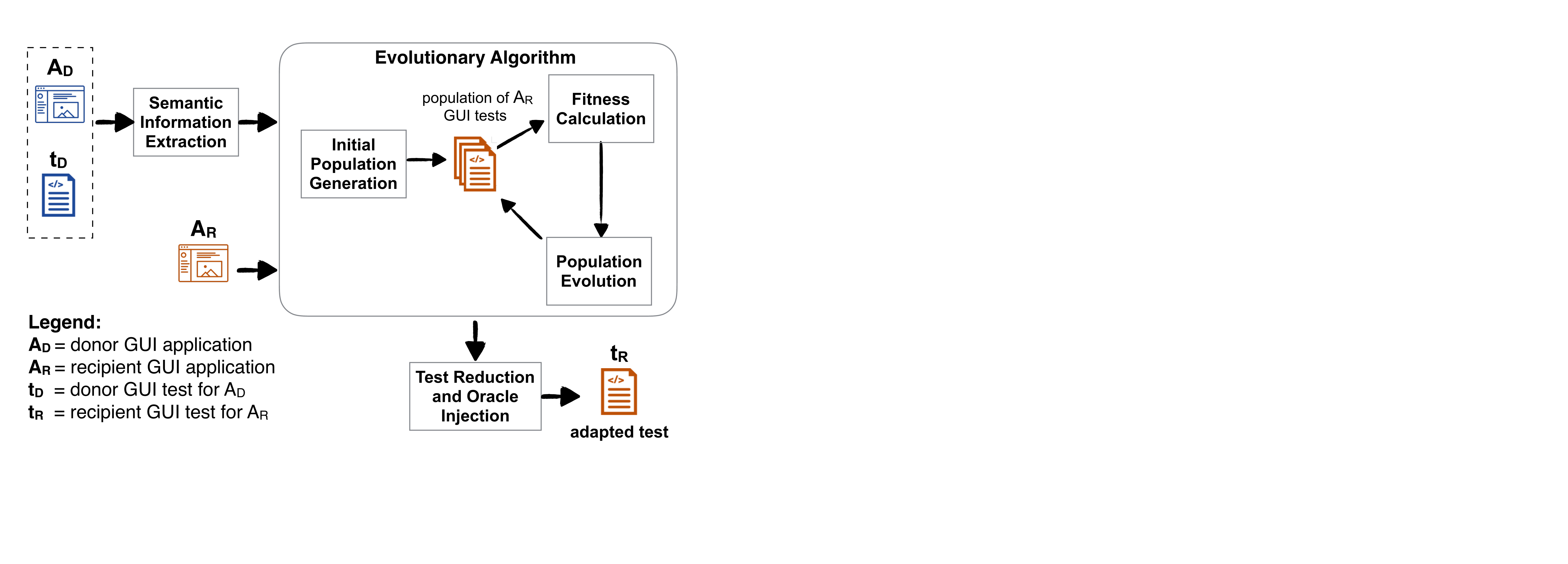}
	}
	\vspace{-8mm}
	\caption{The \nametool  process}
	\label{workflowAdaptdroid}
	
\end{figure}
\label{adaptdroid}

Figure~\ref{workflowAdaptdroid} overviews the \nametool process.
\tool takes as an input the APK of a donor application $A_D$, a donor test $t_D$ and the APK of a recipient application $A_R$, and generates a test  $t_R$ for  $A_R$. 

\tool adapts tests in five phases. 
The \emph{Semantic Information Extraction} phase executes $t_D$ on $A_D$ to extract semantic information relevant to the adaptation process, such as the semantic descriptors of the widgets exercised by $t_D$. The \emph{Initial Population Generation}, \emph{Fitness Calculation} and \emph{Population Evolution} phases implement an evolutionary algorithm that evolves a population of GUI tests guided by a fitness function that steers the evolution towards a test $t_R$ as similar as possible to $t_D$. The \textit{Population Evolution} and \textit{Fitness Calculation} phases iterate until they either perfectly adapt the test (fitness = 1.0) or a time-budget expires. 
The \emph{Test Reduction and Oracle Injection} phase removes irrelevant events in $t_R$ and adds to $t_R$ the oracles adapted from $t_D$.

\nametool faces 
Challenge 1 (huge space of GUI tests) with an \emph{evolutionary algorithm equipped with a proper set of evolution operators}; 
Challenge 2 (GUI differences) with a \emph{matching strategy that takes into account the semantics of GUI events}; 
Challenge 3  (no one-to-one GUI event matching) with a \emph{flexible fitness function} that captures the different nature of the donor and recipient apps.
%
%
The following subsections  describe the cross-app semantic matching of GUI events 
and  the five \nametool phases.

\subsection{Cross-app Semantic Matching of GUI Events} 
\label{sec:semanticMatching}


 \nametool matches GUI events across applications according to their \emph{semantic similarity}, 
regardless of syntactic differences, such as, widget types, positions and layouts.
This is because two similar apps may implement operations that are semantically similar but syntactically different. 

\nametool encodes the semantics of an event as an ordered sequence of one or more words, a sentence in natural language, that we call \emph{descriptor}. 
\tool extracts the event descriptors from either the information shown in the GUI or the identifiers defined by the programmers (widget ids and file names).
Given an event $e_i$, \tool extracts its \textBF{descriptor} $\mathbf{d_i}$ as follows.

For click events $e_i=$~\textit{click$(w_i)$}, $d_{i}$ is the text displayed in the widget~$w_i$
(its text property). In the example of Figure~\ref{adaptExample}, the label of widget $w_{D4}$ \quotes\texttt{Test}\quotes $ $ is the descriptor of $e_{D4}$. Similarly, $d_{R5} = $~\quotes\texttt{Test}\quotes, $d_{R6} = $~\quotes\texttt{Mark as Paid}\quotes  $ $ and $d_{R7} = $ \quotes\texttt{8}\quotes.
If the text property is empty and the widget $w_i$ includes an image, $d_{i}$ is the file name
of the image. 
In the example of Figure~\ref{adaptExample}, the name of the image file associated with $w_{D1}$  (\quotes\texttt{bs\_add\_task}\quotes) is the descriptor of $e_{D1}$. Similarly, $d_{D3} = $  \quotes\texttt{action\_save\_task}\quotes, $d_{R1} = $ 	\quotes\texttt{action\_add}\quotes $ $ and $d_{R4} = $ \quotes\texttt{action\_save}\quotes.
To facilitate the matching of descriptors,  \nametool splits words by underscore or camel-case ($d_{D1}$ becomes \quotes\texttt{bs add task}\quotes), 
removes stop-words, and performs lemmatization~\cite{Manning:CoreNLP:2014}.
If the text property is empty and $w_i$ does not include an image,  $d_{i}$ is the $id$ assigned by the developers to reference $w_i$ in the GUI.

For fill events $e_i = $ \textit{fill($w_i$, \textit{txt}$_i$)}, $d_{i}$ is
the text of the nearest widget from $w_i$.
We follow the approach of Becce et
al.~\cite{becce:descriptors:FASE:2012}, which is based on the observation that text fields
are normally described by near labels.
In the example of Figure~\ref{adaptExample}, the text property of the label on top of $w_{R3}$ \quotes\texttt{Amount}\quotes $ $ is the descriptor of $e_{R3}$.  Similarly, $d_{D2} = $\quotes\texttt{What is to be done?}\quotes and $d_{R2} = $ \quotes\texttt{Payee/Item}\quotes. 
If there are no lables near $w_i$,  $d_{i}$ is the $id$ assigned by the developers to reference $w_i$ in the GUI.

\nametool identifies pairs of descriptors that represent the same concept
with a Boolean function \textBF{\textsc{IsSemSim}\textit{(txt$\mathbf{_1}$,txt$\mathbf{_2}$)}} that returns \T if the sentences \emph{txt$_1$} and  \emph{txt$_2$} are \textBF{semantically similar}, \F otherwise. 
The many available distances, such as Euclidean Distance, Cosine Distance and Jaccard Similarity, are ill-suited for our purposes.
This is because they are not designed to overcome the \emph{synonym problem}, 
that is, two sentences have the same semantics but no common words~\cite{Mikolov:VectorSum:NIPS:2013}.
We cannot expect that two distinct albeit similar apps use exactly the same words to express the same concepts. 

Both \textsc{CraftDroid}~\cite{lin:craftdroid:ASE:2019}  and \textsc{AppTestMigrator}~\cite{behrang:apptestmigrator:ASE:2019} address the synonym problem with \textsc{Word2vec}~\cite{Mikolov:VectorSum:NIPS:2013}, a vector-based \emph{word embedding}~\cite{turian:wordembeddings:2010}. 
\textsc{Word2vec} trains a model that embeds words into a vector space, where words with similar semantics are close in the space~\cite{Mikolov:VectorSum:NIPS:2013}. 
\textsc{Word2vec} matches single words, and thus it is inadequate when descriptors contain multiple words (as $d_{R6} = $~\quotes\texttt{Mark as Paid}\quotes $ $ in Figure~\ref{adaptExample}). 

Instead, \tool uses Word Mover's Distance (WMD)~\cite{Kusner:WMD:ICML:2015}, which calculates the distance between sentences composed of one or more words~\cite{turian:wordembeddings:2010}. 
Given two sentences \textit{txt$_1$} and \textit{txt$_2$}, WMD returns a number between 0 to 1 that expresses how close the sentences are
in the vector space~\cite{Kusner:WMD:ICML:2015}.
 \textsc{IsSemSim}\textit{(txt$_1$, txt$_2$)} = \T, if WMD(txt$_1$, txt$_2$) is greater than a given threshold $\tau$ (0.65 in our experiments), \F otherwise.
 We implement \textsc{IsSemSim} as a Boolean function with a threshold, 
 because the WMD distances are not accurate enough to assume that the highest
similarity is the best one~\cite{Kusner:WMD:ICML:2015,turian:wordembeddings:2010}. 
For example,  two sentences with WMD 0.99 might not be more semantically similar than two sentences with WMD 0.88~\cite{Kusner:WMD:ICML:2015}.

\medskip
We now define the \textbf{semantic matching of events, denoted by~$\mathbf{\sim}$}. Given a donor test $t_D$, a recipient test $t_R$, and two events $e_i \in t_D$ and $e_j \in t_R$, with descriptor $d_i$ and $d_j$, respectively, we say that $\mathbf{e_i  \sim e_j}$ if one of the following cases holds.

\smallskip
\textBF{Matching click events}: $e_i=\textit{click}(w_i) \land e_j =\textit{click}(w_j) \land \textsc{IsSemSim}(d_i,d_j) = $ \T.
This is the case of clicks events that execute a similar functionality.

\smallskip
\textBF{Matching fill events}: $e_i =\textit{fill}(w_i,\textit{txt}_i)\land e_j =\textit{fill}(w_j,\textit{txt}_j) \land \textsc{IsSemSim}(d_i,d_j) = $ \T $ \land $ $ \textit{txt}_i=\textit{txt}_j$. This is the case of fill events 
that execute a similar functionality with the same input. 

\smallskip
\textBF{Matching fill-to-click events}: $e_i=\textit{fill}(w_i,\textit{txt}_i)  \land e_j=\textit{click}(w_j) \land \textsc{IsSemSim}(\textit{txt}_i,d_j)= $ \T.
This is the case of a fill event in $t_D$ that can be mapped to an equivalent click event in $t_R$. For example, entering the value \quotes1\quotes $ $ in a calculator app can be mapped to clicking the button with text \quotes\texttt{1}\quotes~in another calculator app. 
The reader should notice that 
we do not allow the opposite, that is, mapping click events of $t_D$ to fill events of $t_R$. Otherwise, \nametool could easily (and incorrectly) map a button click in $t_D$ with an event of $A_R$ that enters the label of the clicked button in an input field.

In the example of Figure~\ref{adaptExample}, \nametool matches the events in $t_D$ with those in
$t_R$ as follows: 

\noindent
\hspace{6mm} $e_{D1} \sim e_{R1}$ ~~ $e_{D3} \sim e_{R1}$ ~~ $e_{D3} \sim e_{R4}$ ~~
$e_{D4} \sim e_{R5}$  

\label{order}
\subsection{Semantic Information Extraction}


This phase executes $t_D$ in $A_D$ to collect the following information, which are required by the next phases.

- \textBF{Oracle assertions.}  For each oracle assertion in $O_t$, \nametool logs both the state of the widgets, when each assertion is checked, and the expected value of the assertion.

- \textBF{Events ordering.} Obtaining a meaningful test adaptation that preserves the semantics of $t_D$ may require that some events are executed in a specific order. Conversely, 
certain events 
may follow alternative orders without affecting the semantics of the test (such as, the fill events that fill a form). \nametool identifies such events to avoid unnecessary constraints on the events ordering while generating the adapted test.

To identify the opportunity of re-ordering events, \nametool checks if each pair of consecutive events $e_i$ and $e_{i+1}$ in $t_D$ could be potentially executed in the opposite order. Let us consider $\ldots S_{i-1}$ $\xrightarrow{e_i} S_i$ $\xrightarrow{e_{i+1}} S_{i+1} \ldots $, which indicates the sequence of states traversed with the execution of events $e_i$ and $e_{i+1}$. We say that events $e_i$ and $e_{i+1}$ \emph{can be reordered}, denoted by $e_i \rightleftharpoons e_{i+1}$, iff $e_{i+1}$ is enabled in state $S_{i-1}$ and $e_i$ is enabled in state $S_{i+1}$. We say that an event $e$ that interacts with a widget $w$ is enabled in a state $S$ iff $S$ contains a widget $w'$ with the same xpath of the widget $w$ and $w'$ is interactable.

We define the \emph{cluster of the events that can be arbitrarily reordered} as the set of consecutive events that can be reordered. 
Formally, given $e_i \rightleftharpoons e_{i+1}, \forall i=j \ldots m$ $(1 \le j < m < n)$, the corresponding cluster 
is $C=\{e_j, \ldots e_{m+1}\}$. We also say that $\textit{cluster}(e_i)=C, \forall i=j\ldots m+1$. 
We build the clusters by checking each pair of consecutive events to guarantee a linear time complexity with respect to test length. 
For instance, $t_D$ in Figure~\ref{adaptExample} has
four clusters with a single event each, indicating that the prescribed order is the only possible one.

To facilitate the definition of the next phases, we introduce the Boolean function \textBF{\textsc{isBefore}$(e_i, e_j)$} that returns \T iff $cluster(e_i) \neq cluster(e_j) \land i < j$ ($e_i$ must strictly precede $e_j$ in $t_D$), \F otherwise.

\subsection{Initial Population Generation}

Any evolutionary algorithm starts by generating $\mathcal{P}_0$ the initial population of $N$ individuals~\cite{Back:evolutionary:book:1996}. 
An individual for \nametool is a test $t_R$ for the recipient application $A_R$. 
\nametool populates $\mathcal{P}_0$ with $\textit{NR}$ randomly-generated tests (to guarantee
genetic diversity in $\mathcal{P}_0$) and \textit{NG}  tests ($N = \textit{NG}+\textit{NR}$) generated with a greedy algorithm that are similar to $t_D$ 
(to have ``good'' genetic material for evolution).

\tool generates random tests following standard random approaches~\cite{ machiry:dynodroid:FSE:2013}.
Specifically, \tool generates a random test $t_R$ by opening/restarting $A_R$ to obtain the initial state $S_0$, and repeating the following three steps until $t_R$ reaches the maximum length $L$: 
\begin{inparaenum}[(i)] \item it randomly selects an event $e_i$ from those enabled in the current GUI state $S_{i-1}$; \item it appends $e_i$ to $t_R$; \item it executes $e_i$ obtaining the state $S_i$. \end{inparaenum} 

The greedy-algorithm chooses an event $e_i$ among the events that semantically match an event in $t_D$, and then executes step (ii) and (iii) of the random-algorithm. 
In details, step (i) of the greedy-algorithm selects an event $e_i$ from the set $\{e_j ~:~e_j \ \textit{is}\ \textit{enabled}\ \textit{in}\ S_{i-1} \wedge \exists e_k \in t_D : e_j \sim e_k\}$. If this set is empty, it selects an event at random.


\subsection{Fitness Calculation}

\label{fitness}




At each generation \textit{gen} of the evolutionary algorithm, the \textit{Fitness Calculation}  computes a fitness score in [0,1] for each test $t_R$ in $\mathcal{P}_{\textit{gen}}$.
The score characterizes the similarity between $t_R$ and $t_D$, and guides the exploration of possible test adaptations. 
\tool computes the fitness score by executing each $t_R$ in $\mathcal{P}_{\textit{gen}}$ and extracting the event descriptors and state transitions.
While executing the tests, \tool also updates a \emph{GUI model}~\cite{Memon:GUIRipping:WCRE:2003} that encodes the sequence of events that trigger window transitions.
The definition of such a model follows the one proposed by Memon et al.~\cite{Memon:GUIRipping:WCRE:2003}. 
\tool uses this model in the \textit{Population Evolution} phase to repair infeasible tests.
%


We define the fitness function of a test $t_R$, by considering 
\begin{inparaenum}[(i)]
	\item the number of events in $t_D$ that semantically match the events in $t_R$ (\emph{similar events}), and
	\item the number of assertions in $t_D$ that are applicable to the states reached by $t_R$ (\emph{applicable assertions}). Intuitively, the higher these numbers are the more successful the adaptation is.
\end{inparaenum}

\medskip
\noindent
\mbox{\textBF{Similar Events}}
To compute  the number of similar events for each test $t_R \in \mathcal{P}_{\textit{gen}}$, \nametool maps the events in $t_R$ to those in $t_D$ using the semantic matching (see Section~\ref{sec:semanticMatching}).
%
%
Let $\mathcal{M} \subseteq t_R \times t_D$ denote a binary relation over $t_R$ and $t_D$, that we call \textbf{mapping}, such that each pair of events semantically match. That is, $\mathcal{M}$ is a set of pairs of events $(e_R \in t_R, e_D \in t_D) : \forall (e_R,e_D) \in \mathcal{M}, e_R \sim e_D$.
An event in $t_R$ can be mapped to multiple events in $t_D$. 
For instance, in Figure~\ref{adaptExample} $e_{R1}$ maps both $e_{D1}$ and $e_{D3}$. 
We use $\mathbb{M}$ to denote all the possible mappings between events in $t_R$ and $t_D$. 

Many mappings in  $\mathbb{M}$ could be invalid. A mapping $\mathcal{M} \in \mathbb{M}$  is \textbf{valid} iff all the following three criteria are satisfied: 


\smallskip
\noindent
\emph{1) Injective matching}   $\mathcal{M}$ does not contain any event in $t_R$ that relates with more than an event in $t_D$: 
$\forall e_{RA}, e_{RB} \in t_R$ and $\forall e_D \in t_D$, if $(e_{RA},e_D) \in \mathcal{M} \land (e_{RB},e_D) \in \mathcal{M}$, then $RA = RB$ ($e_{RA}$ and $e_{RB}$ are the same event). 
In the example of Figure~\ref{adaptExample}, the mapping $\mathcal{M}=\{(e_{R1},e_{D3}), (e_{R4},e_{D3})\}$  is invalid because it does not satisfy this criterion.

\smallskip
\noindent
\emph{2) Valid ordering} 
All events in $t_R$ satisfy the ordering of $t_D$ as extracted in the \textit{Semantic Information Extraction} phase:  $\forall (e_{RA},e_{DA}),(e_{RB},e_{DB})\in\mathcal{M}$ if \textsc{isBefore}$(e_{DA}, e_{DB})$$=$\T, $RA$$<$$RB$ ($e_{RA}$ precedes $e_{RB}$ in $t_R$).

\smallskip
\noindent
\emph{3) Consistent matching}  
Two events in $t_D$ that are associated with the same event descriptor must be matched to consistent recipient events in $t_R$:
 $\forall (e_{RA},e_{DA}), (e_{RB},e_{DB}) \in \mathcal{M}$ if $e_{DA}$ and  $e_{DB}$ have identical descriptors ($d_{e_{DA}} = d_{e_{DB}}$), then also $e_{RA}$ and  $e_{RB}$ must have identical descriptors ($d_{e_{RA}} = d_{e_{RB}}$).
This constraint avoids mapping two equivalent events in $t_D$ (such as clicking  the same button) to different widgets in $A_R$.


\nametool selects the valid mapping $\mathbf{\mathcal{M}^\star} \in \mathbb{M}$ that maximizes the number of matched events, and uses $\mathbf{\mathcal{M}^\star}$ to compute the event similarity between the two tests. 
%
Intuitively,  $\mathbf{\mathcal{M}^\star}$ is the mapping that best captures the similarity of $t_R$ and $t_D$. 
More formally, $\mathcal{M}^\star \in \mathbb{M}$ such that  $\mathcal{M}^\star$ is valid and $\nexists $ a valid $\mathcal{M} \in \mathbb{M} : \mid$$\mathcal{M}$$\mid > \mid$$\mathcal{M}^\star$$\mid$. 
$\mid$$\mathcal{M}$$\mid$ indicates the number of pairs in a mapping $\mathcal{M}$. 
If there are multiple valid mappings with the highest cardinality, \nametool selects one randomly.
In the example of Figure~\ref{adaptExample}, $\mathcal{M}^\star = \{(e_{R1},e_{D2}), (e_{R4},e_{D4}), (e_{R5},e_{D6})\}$. 

Because of the huge number of possible mappings ($2^{|t_R|\cdot |t_D|}$), \tool does not enumerate $\mathbb{M}$ and then remove all invalid mappings.
Instead, \tool efficiently identifies $\mathcal{M}^\star$ by applying the three validity criteria while building $\mathbb{M}$. 


\medskip
\noindent
\textbf{Applicable Assertions}
\tool fitness function also considers the number of assertions in $t_D$ that ``\emph{can be applied to}'' $t_R$. This is because a good adaptation of the donor test $t_D$ must reach a state of the recipient app with widgets that are compatible with the ones checked by the donor assertions. 

Intuitively, an assertion $o \in O_D$ is \emph{applicable} in $t_R$ if $o$ can be applied to at least a state reached after the execution of the last event in the mapping $\mathcal{M}^\star$ (we recall that we only consider assertions at the end of the tests). 
The applicability of an assertion in a state depends on the existence (or absence) of widgets in the recipient app that are semantically similar to the widgets checked by the donor assertion in the donor app.


 \tool supports four types of assertions: $o_1=$~\textit{exists(txt)} and $o_2=$~\textit{hasText(w,txt)}), and their negative counterparts: $\overline{o_1}=$~\textit{not(exists(txt))} and $\overline{o_2}=$~\textit{not(hasText(w,txt))}.

For the positive assertion types $o_1$ and $o_2$, the Boolean function $\textsc{isApplicable}(o,\mathcal{M}^\star)$ returns \T iff $o$ is \emph{applicable} in the state reached after executing the last event of $t_R$ in $\mathcal{M}^\star$, \F otherwise.
An assertion $o$ is \emph{applicable} in a state $S_i$ if there exists a widget $w^\prime \in S_i$ such that \textsc{isSemSim}$(d_{w^\prime}, d_o) =$~\T, where  $d_{w^\prime}$ is the descriptor of the widget $w^\prime$ extracted with the rules in Section~\ref{sec:semanticMatching}.
The descriptor of an assertion of type $o_1=$~\textit{exists(txt)} is $d_{o1}$ = \textit{txt} , while for type $o_2=$~\textit{hasText($w$,txt)} is $d_{o2} = d_{w}$.

For the negated assertion types $\overline{o_1}$ and  $\overline{o_2}$, we define the $\textsc{isApplicable}$ function differently. 
This is because it is trivial to find a state that does not contain a certain widget/text. Indeed, most
of the states traversed by an adapted test satisfy this condition.
To better capture the semantics of negated assertions, we force $t_R$ to explicitly move the recipient app from a state that does not satisfy the assertion to a state that satisfies it. 
Since we check for the absence of a certain widget/text, we also require $t_R$ to satisfy this constraint on the same window.
Otherwise, the constraint could be easily satisfied by changing the current window of the app. 
More formally, given a negated assertion $\overline{o}$, $\textsc{isApplicable}(\overline{o}, M^\star)$ returns \T iff (i) the positive version of $\overline{o}$ (obtained by removing \textit{not}) is applicable in a state $S_i$ traversed by $t_R$, (ii) the positive version of $\overline{o}$ is not applicable in a state $S_j$ traversed after the last event in the mapping $M^\star$, (iii) $S_i$ is traversed before $S_j$, and (iv) both $S_i$ and $S_j$ refer to the same window; \F otherwise. 

In the example of Figure~\ref{adaptExample}, assertion $o_{D1}$ in $t_D$ verifies that no widget with text \quotes\texttt{Test}\quotes $ $ exists. The assertion is applicable to~$t_R$ because the last state of $t_R$ does not contain such a widget ($o_{D1}$~is~\T), the state after the event $e_{R4}$  does ($o_{D1}$~is~\F), and these two states belong to the same window.

Let $O_D^\star \subseteq O_D$ denote the set of assertions of $t_D$ such that $\textsc{isApplicable}(o,M^\star)$ returns \T. As such, the cardinality of  $O_D^\star$ ($\mid O_D^\star\mid$) measures \emph{the number of assertions successfully adapted to the recipient app}.
%
\setlength{\belowdisplayskip}{10pt}
\setlength{\abovedisplayshortskip}{0pt}
\setlength{\belowdisplayshortskip}{10pt}
\vspace{-0.3mm}
\begin{equation*}
\textsc{\textbf{fitness-score}}\mathbf{(t_R)} =\dfrac{\mid M^\star\mid + \mid O_D^\star \mid}{ \mid t_D\mid + \mid 	O_D \mid} \in [0;1]
\end{equation*}  

The fitness score is proportional to both the number of events and the number of assertions in $t_D$. 
That is, obtaining an applicable assertion contributes as much as successfully adapting an event. 
The score is a value in [0, 1], with 1 representing a perfect adaptation.

\subsection{Population Evolution}

The \textit{Population Evolution} phase combines and mutates the individuals (GUI tests) in the current population $\mathcal{P}_{\textit{gen}}$ to
generate a new population $\mathcal{P}_{\textit{gen+1}}$ of size $N$.
We follow the classic evolutionary algorithm~\cite{whitley:GA:1994}, which works in four consecutive steps: elitism, selection, crossover and mutation.

\smallskip
\noindent
\textBF{Elitism} 
\nametool adds in $\mathcal{P}_{\textit{gen + 1}}$ the elite set  \emph{E} of observed individuals with the highest fitness score ($|E| < N$).
This \emph{elitism} process is a standard genetic algorithm step that avoids missing the best individuals during the evolution~\cite{whitley:GA:1994}. 

\smallskip
\noindent
\textBF{Selection}
\nametool selects $N/2$ pairs of individuals from $\mathcal{P}_{\textit{gen}}$ as candidates for the crossover. 
We use the standard \emph{roulette wheel}~\cite{whitley:GA:1994} selection that assigns at each individual a probability of being selected proportional to its
fitness. 

\smallskip
\noindent
\textbf{Crossover}
\nametool scans each selected pair $\langle t_{R1}$,~$t_{R2} \rangle$ and with probability \textit{CP} performs the crossover and
with probability  $1 - $ \textit{CP} adds the two tests as they are in $\mathcal{P}_{\textit{gen+1}}$. 
The crossover of two parents produces two offspring by swapping their events. \nametool implements a single-point cut crossover~\cite{whitley:GA:1994} as follows.
Given a selected pair $\langle t_{R1}$, $t_{R2} \rangle$, \nametool chooses two random cut points that split both $t_{R1}$ and $t_{R2}$ in two segments. It then creates two new tests. 
One concatenating the first segment of $t_{R1}$ and the second segment of $t_{R2}$. 
The other concatenating the second segment of $t_{R1}$ and the first segment of $t_{R2}$. 

The crossover likely yields infeasible tests, where executing the first segment leads to a window ($W_1$) different from the window ($W_2$) that the first event in the second segment expects. 
\nametool 
repairs these tests
by interleaving the two segments with a sequence of events that move from $W_1$ to $W_2$.
\tool identifies such sequence by querying the GUI Model of $A_R$ (see Section~\ref{fitness}).


\smallskip
\noindent
\textbf{Mutation}
When the crossover terminates ($|P_{\textit{gen+1}}|=N$), \tool mutates the tests in $P_{\textit{gen+1}}$ with a certain probability, aiming to both add genetic diversity and quickly converge to a (sub)optimal solution.  As such, \tool uses two mutations types: random and fitness-driven. 

\emph{Random Mutations} mutate the tests in $P_{\textit{gen+1}}$ with a probability \textit{RM} by applying any of these mutations:
\begin{inparaenum}[(i)] \item adding an event in a random position; \item removing a randomly selected event; \item adding multiple random fill events in a window containing multiple text fields. \end{inparaenum} 
The rationale of the last mutation is that forms with several fields might require many generations to be entirely filled out. This mutation speeds up the evolution by filling all the text fields in a single mutation. 

\emph{Fitness-Driven Mutations} mutate a test to improve its fitness score. Each test in $P_{\textit{gen+1}}$  has a probability \textit{FM} of being mutated using one of these two mutations:
\begin{inparaenum}[(i)] 
 \item  removing an event in $t_R$ that does not match (according to $\mathcal{M}^\star$) any event in $t_D$;
\item adding an event $e_j$ in $t_R$ such that $e_i \sim e_j$, where $e_i$ is a randomly selected event in $t_D$ that does not match $t_R$ events.
\end{inparaenum}



Like crossovers, also mutations could create infeasible tests. 
\nametool identifies them by checking if all the events in the mutated tests can be executed in the order prescribed by the test, and fixes the infeasible tests it by removing all non-executable events. 
Indeed, the fixed test could still have useful genetic material for the evolution~\cite{Harman:SBSE:IST:01}.

The search for an adapted test keeps evolving and evaluating populations of tests until either a predefined budget expires (\# of generations or  time) or \tool finds a test with fitness one. 
When the search terminates, \tool post-processes the test with the highest fitness score by reducing the test length, and injecting the donor assertions (if possible).

 \nametool reduces the test length by removing one by one the events that are not part of the mapping $\mathcal{M}^\star$ used to calculate the fitness score. 
 After removing an event, \nametool executes the test and recalculates its fitness.
  If the fitness decreases, \nametool restores the event because,
even though it did not directly contribute to the fitness value, it
enabled other relevant events to be executed.
In the $\mathcal{M}^\star$ of the example of Figure~\ref{adaptExample}, events $e_{R2}$, $e_{R3}$, $e_{R6}$, and $e_{R7}$ of $t_R$ do not match any event in $t_D$, but the post-process keeps them because removing any event reduces the fitness.

If the fitness function finds some assertions in $t_D$ that are applicable to $t_R$, \nametool adds them at the end of $t_R$. In the example of Figure~\ref{adaptExample}, \tool injects the assertion $o_{D1}= $ \textit{not(exists(\quotes Test\quotes))} at the end of $t_R$. 

\section{Evaluation}
\label{empirical}

We evaluated \nametool by implementing a prototype tool for Android apps~\cite{adaptdroid}. Our prototype uses the \textsc{Appium} 6.1.0 framework~\cite{appium} to read the GUI states
and Android emulators to execute the tests. We evaluated \tool considering two research questions:

\begin{description}
\item[RQ1:] \mbox{\textbf{Effectiveness}} \emph{Can \nametool effectively adapt GUI tests and oracles across similar applications?}
\item[RQ2:] \mbox{\textbf{Baseline Comparison}} \emph{Is \nametool more effective than baseline approaches?}
\end{description}


To measure the quality of test adaptations we need human judgment, possibly involving the designers of the donor tests. 
For this reason, we evaluated \nametool with a human study that involved four PhD students majoring in Software Engineering, who were not related to this project. 
We asked each participant to design some donor tests and to evaluate the adaptations produced by \tool.


\begin{table*}[t]

	\centering
	\caption{Evaluation subjects and results}
		\vspace{-1mm}
	\renewcommand*{\arraystretch}{1.03}
	\setlength{\tabcolsep}{2.7pt}
	\label{table:results}
	\resizebox{1\textwidth}{!}{%
					\rowcolors{1}{}{gray!10}
		\begin{tabular}{ccrlrc|rrr|rrrrc|rrrr}
			\toprule
			\hiderowcolors
			\multicolumn{6}{c|}{\textbf{Subject description}} & \multicolumn{3}{c|}{\textbf{\tool}}            & \multicolumn{5}{c|}{\textbf{RQ1: Effectiveness}}                                                                & \multicolumn{4}{c}{\textbf{RQ2: Baseline}}                                \\
			
			\textbf{Tester} & \textbf{Donor App ($\mathbf{A_D}$})&$\mathbf{\mid}$$\mathbf{t_D}$$\mathbf{\mid}$
			& \textbf{Recipient App ($\mathbf{A_R}$)}&
			\textbf{ID}         & \textbf{$\mathbf{\langle A_D, A_R \rangle}$}        & \textbf{$\mathbf{\mid t_R \mid}$} & \textbf{fitness} & \textbf{gen.} & \textbf{$\mathbf{Q_T}$} & \textbf{\# spurious} & \textbf{\# missing} & \textbf{$\mathbf{Q_S}$} & \textbf{Oracle} & \multicolumn{2}{c}{\textbf{Random}} &  \multicolumn{2}{c}{\textbf{Basic}} \vspace{-0.5mm}
			\\ 
			& & 
			&&
			&    \textbf{adaptable?}  & & \textbf{} & \textbf{} & \textbf{} & \textbf{events} & \textbf{events} & \textbf{} & \textbf{adapted?} & \textbf{fitness} & \textbf{gen.} & \textbf{fitness} & \textbf{gen.} \\ \midrule
			
	\multirow{6}{*}{T1} & \multirow{3}{*}{\begin{tabular}[c]{@{}c@{}}\href{https://play.google.com/store/apps/details?id=at.markushi.expensemanager}{Expense Manager}\\ (Expense Tracking)\end{tabular}} & \multirow{3}{*}{15} &\cellcolor{gray!10}\href{https://play.google.com/store/apps/details?id=com.kpmoney.android}{KPmoney} & 		\cellcolor{gray!10} 1                   &		\cellcolor{gray!10}Partially                  &		\cellcolor{gray!10} 13                                &		\cellcolor{gray!10} \textbf{0.64}                   & 		\cellcolor{gray!10}33           &		\cellcolor{gray!10} 3                       &		\cellcolor{gray!10} 6                   &		\cellcolor{gray!10} 0                   & 		\cellcolor{gray!10}1.00                   &		\cellcolor{gray!10}No              &		\cellcolor{gray!10} 0.30                   &		\cellcolor{gray!10} 79           & 0.57                   & 		\cellcolor{gray!10}50           \\
			
			&  &  &  \href{https://play.google.com/store/apps/details?id=com.monefy.app.lite}{Monefy} & 2                    &Yes                        & 10                                & \textbf{0.47}                   & 59           & 4                       & 1                   & 0                   & 1.00                   &Yes             & 0.23                   & 3            & 0.43                   & 98           \\
			
			&  &  &  \cellcolor{gray!10}\href{https://play.google.com/store/apps/details?id=com.realbyteapps.moneymanagerfree}{Money} & \cellcolor{gray!10}  3                   & \cellcolor{gray!10}Yes                        & \cellcolor{gray!10}  15                                & \cellcolor{gray!10}  \textbf{0.47}                   & \cellcolor{gray!10}  95           & \cellcolor{gray!10}  4                       & \cellcolor{gray!10}  0                   & \cellcolor{gray!10}  0                   & \cellcolor{gray!10}  1.00                   & \cellcolor{gray!10}  Partially       & \cellcolor{gray!10}  0.30                   & \cellcolor{gray!10}  2            & \cellcolor{gray!10}  0.40                    & \cellcolor{gray!10}  82       
			\\ \cline{2-18} 
			
			& \multirow{3}{*}{\begin{tabular}[c]{@{}c@{}}\href{https://play.google.com/store/apps/details?id=com.mirte.notebook}{Mirte Notebook} \\ (Note Keeping)\end{tabular}} & \multirow{3}{*}{10} &  \href{https://play.google.com/store/apps/details?id=com.hlcsdev.x.notepad}{Xnotepad} & 4                   & Partially                  & 7                                 & \textbf{0.78}                   & 91           & 1                       & 4                   & 3                   & 0.50                   & No              & 0.36                   & 8            & 0.77                   & 80           \\
			
			&  &   & \cellcolor{gray!10}\href{https://play.google.com/store/apps/details?id=com.socialnmobile.dictapps.notepad.color.note}{Color Notes} & \cellcolor{gray!10}  5                   & \cellcolor{gray!10}Yes                        & \cellcolor{gray!10}  15                                & \cellcolor{gray!10}0.24                   & \cellcolor{gray!10}  4            & \cellcolor{gray!10}  2                       & \cellcolor{gray!10}  9                   & \cellcolor{gray!10}  2                   & \cellcolor{gray!10}  0.75                  & \cellcolor{gray!10}Partially       & \cellcolor{gray!10}  0.21                   & \cellcolor{gray!10}  40           & \cellcolor{gray!10}  \textbf{0.27}                   & \cellcolor{gray!10}  9            \\
			
			&  &  & \href{https://play.google.com/store/apps/details?id=org.whiteglow.keepmynotes}{Keep Mynotes} & 6                   & Yes                        & 2                                 & \textbf{0.39}                   & 1            & 1                       & 1                   & 5                   & 0.14                   & No              & 0.33                   & 79           & 0.33                   & 11           \\ \hline
			
			\multirow{6}{*}{T2} & \multirow{3}{*}{\begin{tabular}[c]{@{}c@{}}\href{https://play.google.com/store/apps/details?id=at.markushi.expensemanager}{Markushi Manager}\\ (Expense Tracking)\end{tabular}} & \multirow{3}{*}{16} & \cellcolor{gray!10}\href{https://play.google.com/store/apps/details?id=com.mhriley.spendingtracker}{Spending Tracker} & \cellcolor{gray!10}  7                   & \cellcolor{gray!10}Yes                        & \cellcolor{gray!10}  23                                & \cellcolor{gray!10}  \textbf{0.74}                   & \cellcolor{gray!10}  59           & \cellcolor{gray!10}  2                       & \cellcolor{gray!10}  1                   & \cellcolor{gray!10}  8                   & \cellcolor{gray!10}  0.73                   & \cellcolor{gray!10}Partially       & \cellcolor{gray!10}  0.72                   & \cellcolor{gray!10}  7            & \cellcolor{gray!10}  \textbf{0.74}                   & \cellcolor{gray!10}  36           \\

			&  &  & \href{https://play.google.com/store/apps/details?id=com.smartexpenditure}{Smart Expenditure} & 8                   & Yes                        & 18                                & \textbf{0.41}                   & 25           & 0                       & -                   & -                   & -                      & -             & 0.32                   & 3            & \textbf{0.41}                   & 16           \\
			
			&  &  & \cellcolor{gray!10}\href{https://play.google.com/store/apps/details?id=mic.app.gastosdiarios_clasico}{Gastos Diarios} & \cellcolor{gray!10} 9                   & \cellcolor{gray!10}Partially                  & \cellcolor{gray!10} 6                                 & \cellcolor{gray!10} \textbf{0.50}                    & \cellcolor{gray!10} 17           & \cellcolor{gray!10} 0                       & \cellcolor{gray!10} -                   & \cellcolor{gray!10} -                   & \cellcolor{gray!10} -                      & \cellcolor{gray!10} -             & \cellcolor{gray!10} 0.37                   & \cellcolor{gray!10} 19           & \cellcolor{gray!10} \textbf{0.50}                    & \cellcolor{gray!10} 22           \\ \cline{2-18} 
			
			& \multirow{3}{*}{\begin{tabular}[c]{@{}c@{}}\href{https://play.google.com/store/apps/details?id=com.bitslate.notebook}{Bitslate Notebook} \\ (Note Keeping)\end{tabular}} & \multirow{3}{*}{13} &  \href{https://play.google.com/store/apps/details?id=com.studio.tools.one.a.notes}{Notes} & 10                  & Yes                        & 8                                 & \textbf{0.45}                   & 15           & 3                       & 0                   & 2                   & 0.80                   & No              & 0.33                   & 61           & 0.31                   & 10           \\

			&  &  &\cellcolor{gray!10}\href{https://play.google.com/store/apps/details?id=com.taxaly.noteme.v2}{Fast Notepad} &\cellcolor{gray!10} 11                  &\cellcolor{gray!10}Yes                        &\cellcolor{gray!10} 10                                &\cellcolor{gray!10} 0.41                   &\cellcolor{gray!10} 86           &\cellcolor{gray!10} 4                       &\cellcolor{gray!10} 0                   &\cellcolor{gray!10} 2                   &\cellcolor{gray!10} 0.83                   &\cellcolor{gray!10}Yes             &\cellcolor{gray!10} 0.23                   &\cellcolor{gray!10} 6            &\cellcolor{gray!10} \textbf{0.44}                  &\cellcolor{gray!10} 41           \\
			
			&  &  &  \href{https://play.google.com/store/apps/details?id=ru.andrey.notepad}{Notepad} & 12                  & Yes                        & 12                                & \textbf{0.11} & 1            & 1                       & 10                  & 11                  & 0.15                  & Yes             & \textbf{0.11}                   & 1            & \textbf{0.11}                   & 1            \\ \hline
			
			\multirow{6}{*}{T3} & \multirow{3}{*}{\begin{tabular}[c]{@{}c@{}}\href{https://play.google.com/store/apps/details?id=com.pocketuniverse.ike}{Pocket Universe} \\ (To-dolist)\end{tabular}} & \multirow{3}{*}{11}  &\cellcolor{gray!10}\href{https://play.google.com/store/apps/details?id=org.whiteglow.keepmynotes}{Seven Habits} &\cellcolor{gray!10} 13                  &\cellcolor{gray!10}No                         &\cellcolor{gray!10} -                                 &\cellcolor{gray!10}             -           &\cellcolor{gray!10}     -         &\cellcolor{gray!10} -                       &\cellcolor{gray!10} -                   &\cellcolor{gray!10} -                   &\cellcolor{gray!10}          -              &\cellcolor{gray!10} -               &\cellcolor{gray!10}              -          &\cellcolor{gray!10}        -      &\cellcolor{gray!10}            -            &\cellcolor{gray!10}       -       \\
			
			&  &  & \href{https://play.google.com/store/apps/details?id=com.obplanner}{Ob Planner} & 14                  & No                         & -                                 &     -                   &       -       & -                       & -                   & -                   &         -               & -               &          -              &      -        &          -              &      -        \\
			
			&  &  &\cellcolor{gray!10}\href{https://play.google.com/store/apps/details?id=jakiganicsystems.simplestchecklist}{Simplest Checklist} &\cellcolor{gray!10} 15                  &\cellcolor{gray!10}No                         &\cellcolor{gray!10} -                                 &\cellcolor{gray!10}       -                 &\cellcolor{gray!10}       -       &\cellcolor{gray!10} -                       &\cellcolor{gray!10} -                   &\cellcolor{gray!10} -                   &\cellcolor{gray!10}         -               &\cellcolor{gray!10} -               &\cellcolor{gray!10}        -                &\cellcolor{gray!10}     -         &\cellcolor{gray!10}          -              &\cellcolor{gray!10}     -         \\ \cline{2-18} 
			& \multirow{3}{*}{\begin{tabular}[c]{@{}c@{}}\href{https://play.google.com/store/apps/details?id=com.alibaba.aliexpresshd}{Aliexpress}\\ (Online Shopping)\end{tabular}} & \multirow{3}{*}{16} &  \href{https://play.google.com/store/apps/details?id=com.banggood.client}{Banggood}& 16                  & Yes                        & 11                                & 0.50                   & 43           & 2                       & 4                   & 1                   & 0.88                   & No              & 0.30                   & 11           & \textbf{0.64}                   & 36           \\
			
			&  &  &\cellcolor{gray!10}\href{https://play.google.com/store/apps/details?id=com.lightinthebox.android}{Light in the box} &\cellcolor{gray!10} 17                  &\cellcolor{gray!10}Yes                        &\cellcolor{gray!10} 7                                 &\cellcolor{gray!10} \textbf{0.50}                   &\cellcolor{gray!10} 75           &\cellcolor{gray!10} 1                       &\cellcolor{gray!10} 4                   &\cellcolor{gray!10} 5                   &\cellcolor{gray!10} 0.38                  &\cellcolor{gray!10}No              &\cellcolor{gray!10} 0.23                   &\cellcolor{gray!10} 23           &\cellcolor{gray!10} 0.44                   &\cellcolor{gray!10} 72           \\
			
			&  &  &  \href{https://play.google.com/store/apps/details?id=com.zzkko}{Shein} & 18                  &Yes                        & 9                                 & \textbf{0.30}                   & 92           & 0                       & -                   & -                   & -                      & No              & 0.17                   & 62           & \textbf{0.30}                    & 64           \\\hline
			\multirow{6}{*}{T4} & \multirow{3}{*}{\begin{tabular}[c]{@{}c@{}}\href{https://play.google.com/store/apps/details?id=de.zalando.mobile}{Zalando}  \\ (Online Shopping)\end{tabular}} & \multirow{3}{*}{6} &\cellcolor{gray!10}\href{https://play.google.com/store/apps/details?id=com.inditex.zara}{Zara}  &\cellcolor{gray!10} 19                  &\cellcolor{gray!10}Yes                        &\cellcolor{gray!10} 8                                 &\cellcolor{gray!10} \textbf{0.42}                   &\cellcolor{gray!10} 19           &\cellcolor{gray!10} 3                       &\cellcolor{gray!10} 0                   &\cellcolor{gray!10} 0                   &\cellcolor{gray!10} 1.00                   &\cellcolor{gray!10}No              &\cellcolor{gray!10} 0.38                   &\cellcolor{gray!10} 32           &\cellcolor{gray!10} \textbf{0.42}                   &\cellcolor{gray!10} 22           \\
			&  &  & \href{https://play.google.com/store/apps/details?id=com.romwe}{Romwe} & 20                  & Yes                        & 5                                 & \textbf{0.54}                   & 5            & 3                       & 0                   & 1                   & 0.83                   & No              & 0.50                   & 1            & \textbf{0.54}                   & 11           \\
			&  &  &\cellcolor{gray!10}\href{https://play.google.com/store/apps/details?id=com.yoox}{Yoox} &\cellcolor{gray!10} 21                  &\cellcolor{gray!10}Yes                        &\cellcolor{gray!10} -                                 &\cellcolor{gray!10}         -               &\cellcolor{gray!10}              &\cellcolor{gray!10} -                       &\cellcolor{gray!10} -                   &\cellcolor{gray!10} -                   &\cellcolor{gray!10}   -                     &\cellcolor{gray!10} -               &\cellcolor{gray!10}        -                &\cellcolor{gray!10}       -       &\cellcolor{gray!10}               -         &\cellcolor{gray!10}        -      \\
			\cline{2-18} 
			& \multirow{3}{*}{\begin{tabular}[c]{@{}c@{}}\href{https://play.google.com/store/apps/details?id=com.splendapps.splendo}{Splendo} \\ (To-dolist)\end{tabular}} & \multirow{3}{*}{10} &  \href{https://play.google.com/store/apps/details?id=com.mykhailovdovchenko.to_dolist}{To Do List} & 22                  & Yes                        & 9                                 & \textbf{0.63}                   & 59           & 3                       & 4                   & 6                   & 0.45                   & Yes             & 0.38                   & 46           & 0.55                   & 26           \\
			&  &  &\cellcolor{gray!10}\href{https://play.google.com/store/apps/details?id=com.tasks.android}{Tasks} &\cellcolor{gray!10} 23                  &\cellcolor{gray!10}Yes                        &\cellcolor{gray!10} 8                                 &\cellcolor{gray!10} \textbf{0.46}                   &\cellcolor{gray!10} 32           &\cellcolor{gray!10} 1                       &\cellcolor{gray!10} 6                   &\cellcolor{gray!10} 5                   &\cellcolor{gray!10} 0.29                   &\cellcolor{gray!10}No              &\cellcolor{gray!10} 0.34                   &\cellcolor{gray!10} 59           &\cellcolor{gray!10} 0.38                   &\cellcolor{gray!10} 10           \\
			
			&  &  & \href{https://play.google.com/store/apps/details?id=com.ticktick.task}{Tick Tick} & 24                  & Yes                        & 15                                & 0.46                   & 21           & 1                       & 11                  & 8                   & 0.33                   & Yes             & 0.29                   & 7            & \textbf{0.53}                   & 93 \\
			\midrule
		\end{tabular}

	}
\vspace{-3mm}
\end{table*}

\smallskip
\noindent
\mbox{\textbf{Selecting Subjects and Collecting Donor Tests}}
We selected a total of 32 Android apps (8 donors and 24 recipients) from the Google Play Store by referring to four app categories that represent apps with recurrent functionalities~\cite{Hu:appflow:FSE:2018}: \textit{Expense Tracking}, \textit{To-Do List}, \textit{Note Keeping}, and \textit{Online Shopping}. We avoided selection biases as follows.

We queried the Google Play Store by searching for each category name.
From the list of returned apps, we selected the first two that are  free/freemium and do not require login credentials at start-up.
Thus, obtaining a total of eight donor apps. 
For each donor app $A_D$, we identified three recipient apps by retrieving the list of similar apps suggested in the Google Play Web page of $A_D$.
From this list, we selected the first three apps that were not selected as donors and have the same characteristics described above.
This process resulted in 24 pairs $\langle A_D$, $A_R \rangle$ of donor and recipient apps.

We randomly partitioned the eight donor apps among the testers, by assigning two donor apps of different categories to each tester.
In this way, we prevented that a tester could design similar donor tests.
We asked each tester to design a Selenium GUI test~\cite{selenium} (with an oracle assertion) to test the main functionality of the app.
We left up to the tester to identify the main functionality of the app.

After each tester implemented a donor test, we asked to evaluate whether the test could be adapted to the recipient apps. Each tester evaluated each adaptation on a scale  \quotes Fully\quotes~(the main functionality of $A_R$ can be tested as in $t_D$),  \quotes Partially\quotes~($A_R$ allows to replicate only some of the operations performed in $t_D$),  \quotes No\quotes~($A_R$ implements no functionality that can be tested as in $t_D$). Column ``\emph{$\langle A_D, A_R \rangle$  Adaptable?}'' of Table~\ref{table:results} reports the responses. The testers deemed fully adaptable 18 pairs of tests (75\%) and partially 3 pairs of tests (12.5\%). This result confirms the intuition that GUI tests can be adapted across similar applications. Tester $T_3$ deemed the pairs with ID 13, 14 and 15 as not adaptable, because the test executes functionalities available only in the donor \textit{Pocket Universe} and not in the recipient apps. 
We asked the four testers to manually adapt the fully and partially adaptable donor tests to the recipient apps.

\begin{table}[t]

	\centering
	
	\caption{Configuration parameters of \tool}

	\label{table:config}
	\renewcommand*{\arraystretch}{1}
	\setlength{\tabcolsep}{2pt}
	\resizebox{1\linewidth}{!}{%
		\rowcolors{1}{}{gray!10}
		\begin{tabular}{llr|llr}
			\hiderowcolors
			\toprule
			\textbf{name}      &\textbf{description}                         &\textbf{value} & 			\textbf{name}      &\textbf{description}                         &\textbf{value}\\ \midrule
			\showrowcolors
			$\tau$    & threshold for WMD & 0.65     &
			$N$       & population size                        & 100    \\
			$E$       & \# tests for elitism                   & 10        &
			$L$       & max length of the initial tests        & $|t_D|$   \\
			$\textit{NR}$      & \# initial random tests                & 90      &
			$\textit{NG}$      & \# initial greedy tests                & 10     \\
			$\textit{CP}$       & crossover prob.                  & 0.40       &
			$\textit{RM}$      & random mutation prob.            & 0.35      \\
			$\textit{FM}$      & fitness-driven mutation prob.           & 0.35 & &  &\\
			\bottomrule   
	\end{tabular} }
\end{table}

\smallskip
\noindent
\mbox{\textbf{Running \nametool}}
We ran \nametool with the 21 fully and partially adaptable donor tests 
giving as input the pairs $\langle A_D, A_R\rangle$ and the corresponding
manually-written test $t_D$. 
We used a popular \textsc{WMD} model trained on a \textit{Google News} dataset (about 100 billion words)~\cite{w2vecmodels}. 

We ran \tool with a budget of 100 generations 
with the configuration parameters values shown in Table~\ref{table:config}.
We selected these values by performing some trial runs and by following basic guidelines of genetic programming~\cite{Back:evolutionary:book:1996}.  
Special considerations can be made for the values $\tau$ and $L$. 
We chose $\tau = 0.65$ as the threshold for the semantic similarity by evaluating the WMD model on a list of $\sim$2.5 M synonyms~\cite{moby}. 
More specifically, $\tau = 0.65$ is the threshold that achieves the best trade-off between matched synonyms and unmatched pair of randomly selected words.
We choose $L = |t_D|$ to obtain initial tests for $A_R$ with a max length proportional to the length of the donor test.

When dealing with the test pair with ID 21, we experienced some compatibility issues between the \textsc{Appium} framework and the recipient app $A_R$,  
issues that prevented \tool  generating tests.
Thus, we exclude such a pair from our analysis.

The ``\tool'' columns of Table~\ref{table:results} show information about the returned adapted test $t_R$ (the one with the highest fitness score).
Column ``$|t_R|$'' provides the number of events of the adapted test. Column ``\emph{fitness}'' shows the fitness score of $t_R$. 
\tool never reached fitness score 1.0, thus it terminated after 100 generations.
Column  ``\emph{gen.}'' shows the generation in which \tool produced $t_R$. 

\tool completed 100 generations in 24 hours on average, and spent most of this time
in executing the generated tests on the emulator. 
Executing tests is expensive because \nametool re-installs $A_R$ in the emulator before each test execution to guarantee that each test executes from a clean state. This time could be reduced by running many emulators in parallel or using cloud platforms for mobile testing.

\medskip
\noindent
\mbox{\textbf{RQ1: Effectiveness}}
We asked the testers to judge the quality of each test case $t_R$ produced with \tool for their assigned pairs.
We used a score from 0 to 4, where 0 means that $t_R$ is completely unrelated to the donor test semantics, and 4 means that $t_R$ is an adaptation as good as the one that they manually produced (Column ``\emph{$Q_T$}'' of Table~\ref{table:results}).

In 8 cases out of 20 (40\%) the testers evaluated \tool adaptations as high quality ($Q_T \geq 3 $), with three of which considered perfect adaptations. 
In three cases (15\%), the adapted tests were evaluated as medium quality ($Q_T
= 2$). 
This suggests that in these eleven cases ($Q_T \geq 2 $) the fitness function well describes the similarity with the donor test.

We asked the testers to indicate the spurious and missing events in the tests.  Columns ``$\#$~\textit{spurious events}'' and  ``$\#$~\textit{missing events}'' of Table~\ref{table:results} report the number of events identified as spurious and missing to obtain a perfect adaptation, respectively. 
Column ``$Q_S$'' of Table~\ref{table:results} reports a \emph{structural} quality indicator of the completeness of the matched events:
$Q_S = 1 - (\nicefrac{\# \textit{missing}}{\mid \overline{t_R} \mid})$, where $\overline{t_R}$  is the manually adapted test. 
$Q_S$ ranges in [0, 1], where 0 indicates \emph{no} matching between the events in $t_R$ and $\overline{t_R}$, and 1 indicates a perfect matching.
The average of $Q_S$ is f 0.53, indicating that overall \tool adapted 53\% \emph{true} event matches identified by the testers. 
There is a moderate correlation between the two quality indicators $Q_T$ and $Q_S$ (Pearson coefficient is 0.89). This confirms that it is important for the testers to adapt a large portion of a test.

Column ``\emph{Oracle Adapted?}'' in Table~\ref{table:results} reports whether $t_R$ has an assertion that the tester evaluated to be correct (\quotes Yes\quotes),  partially correct (\quotes Partially\quotes), or not applicable in the states reached by $t_R$ (\quotes No\quotes). 
Testers $T_1$ and $T_2$ attributed partial correctness to three adapted oracles because of marginal differences in the expected output. For instance, in the pair with ID 3, the oracle in the donor test checks if a widget with descriptor \quotes\texttt{expenses}\quotes~has text \quotes\texttt{100}\quotes. The corresponding widget in the recipient test has text \quotes\texttt{-100}\quotes, which is semantically 
equivalent (100 expenses = -100 balance)
but syntactically different. 
Therefore, the oracle was deemed partially correct.

We identify two main issues that limited the effectiveness of \tool.

\smallskip
\emph{1) Significant differences between $A_D$ and $A_R$.}
For instance, in the pair with ID 18, $t_D$ searches in the \emph{Aliexpress} app for a USB drive and adds it to the shopping basket.
 Tester $T_3$ manually adapted $t_D$ searching in the \emph{Shein} app  for a t-shirt.
\nametool failed to adapt this test to \emph{Shein}, as searching for a USB drive in \emph{Shein} results in an empty search. 
As another example, in the pairs with ID 23 and 24, $t_D$ adds a task to a pre-existing \emph{work} task list. 
 \nametool failed to adapt these tests, as the recipient apps do not have a task list. 

\emph{2) Missed Event Matches}. The semantic matching of the descriptors was not always precise due to (i) the
unsoundness of WMD; and (ii) the limited semantics information of the event descriptors.
For example, some of the considered apps have image buttons with file name \quotes\texttt{fabButton.png}\quotes, which 
does not describe the semantics of the widget. 

The results of our study are promising: \tool produced eleven good quality test adaptations between apps with very different GUIs.
In the experiments, we configure \tool to report all adaptations.
We can improve the quality of the generated output, by reporting only adapted tests that reach a minimum fitness score. 

\smallskip
\noindent
\mbox{\textbf{ RQ2: Baseline Comparison}}
We compare \tool with two baseline approaches
\begin{inparaenum}[(i)]
	\item \textsc{Random}, to empirically assess the effectiveness of the evolutionary algorithm of \tool;
	\item \textsc{\tool-basic}, a restrained version of \tool without the fitness-driven mutations and greedy-matching initialization, to assess their impact to the overall effectiveness. \end{inparaenum}

We obtained \textsc{Random} from \tool by
\begin{inparaenum}[(i)]
	\item replacing the roulette-wheel selection with random selection,
	\item  randomly generating the tests in the initial population (\textit{NR} $=100$ and \textit{NG} $= 0$, Table~\ref{table:config}),
	\item setting the probability of the fitness-driven mutations to zero (\textit{FM}$ = 0.0$ Table~\ref{table:config}), and
	\item disabling elitism. 
\end{inparaenum}
As such, \textsc{Random} carries population initialization and evolution completely random. 
We opted to use a random variant of \nametool rather than an existing random generator~\cite{ machiry:dynodroid:FSE:2013}, for a meaningful evaluation. With an existing random generator, we cannot ensure that the differences are due to the search strategy and not to differences in other implementation details, such as the events and inputs types considered by the tools.

We obtained \textsc{\tool-basic} from \tool by applying only the modifications (ii) and (iii) described above.


We ran \textsc{Random} and  \textsc{\tool-basic} with a budget of 100 generations as \nametool. The last four columns in Table~\ref{table:results} show the fitness score of the fittest test and the generation that created it (with the highest fitness value among the tools in bold).
\tool consistently achieves a higher fitness than \textsc{Random}, and the same fitness only in one case (ID 22).
The difference between the tools is statistically significant: a two-tailed t-test returns a p-value of 0.0002. 
\nametool achieves an average fitness of 0.48, while \textsc{\nametool-basic}  of 0.45. This shows a difference albeit small of the fitness score.

\begin{figure}[t!]
	\centering
	\resizebox{\linewidth}{!}{%
		\includegraphics{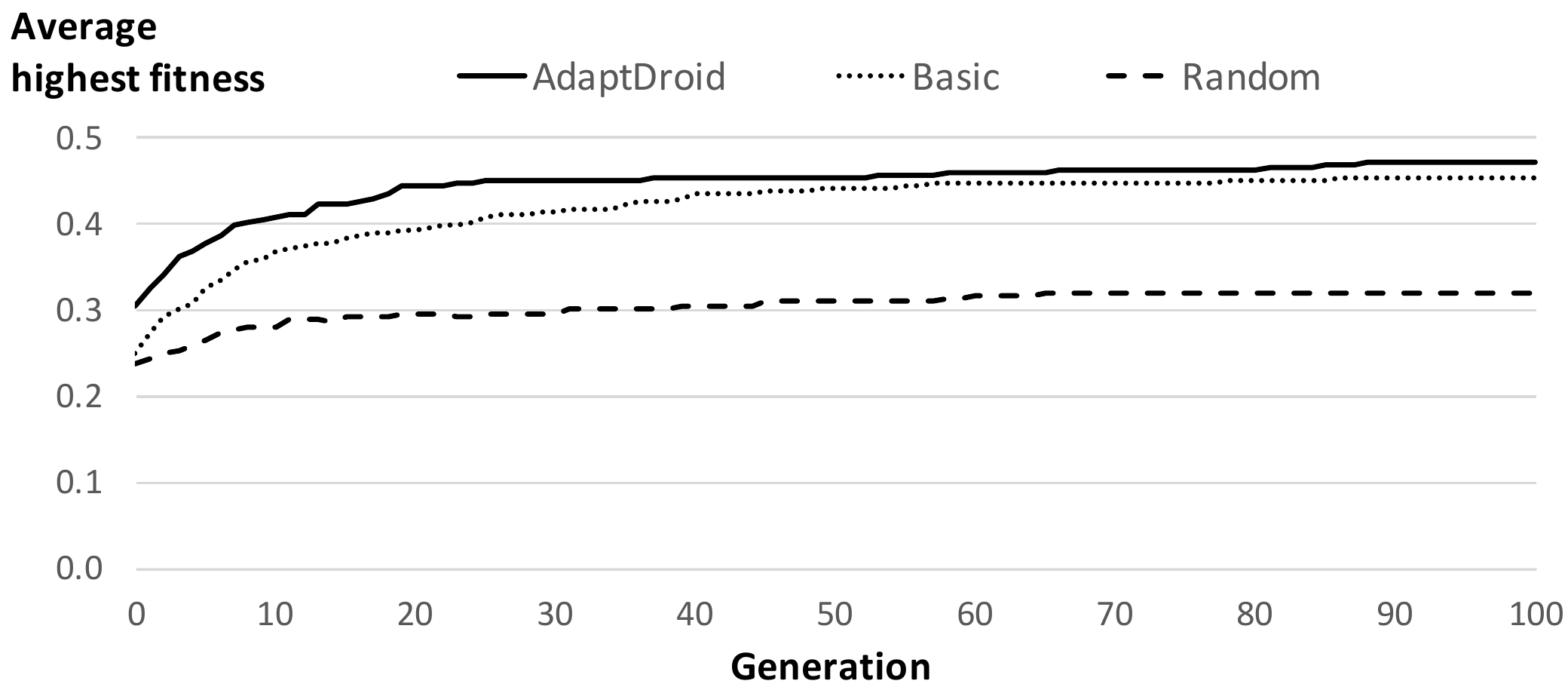}
	}
	\vspace{-7mm}
	\caption{Average fitness growth.}
	\label{fig:fitness}

\end{figure}

Figure~\ref{fig:fitness}  illustrates the gain of \tool over the baseline approaches, by plotting the highest fitness score 
per generation averaged over the 20 adaptations. The plot highlights three important aspects:

\smallskip
\noindent
\textbf{I.-} the fittest test of the greedily generated initial population ($\mathcal{P}_0$) of \tool has a much lower fitness score than the final score. 
This shows that the greedy algorithm used to initialize the population is inadequate, thus motivating the use of an evolutionary approach.

\smallskip
\noindent
\textbf{II.-} 
the highest fitness per generation of \nametool steadily increases while the
fitness of \textsc{Random} saturates much faster. 
This demonstrates that \nametool evolutionary approach is effective in exploring the search space, and it confirms our hypothesis that \nametool generates GUI tests that can hardly be generated at random.

\smallskip
\noindent
\textbf{III.-} 
\nametool and \textsc{\nametool-basic} reach similar fitness, but \nametool
reaches it faster.
This indicates that the greedy-match initialization and the fitness-driven
mutations help to converge faster to the fittest test.

\smallskip
\noindent
\textbf{Threats to Validity} A main threat to the external validity concerns the generalizabilty of a small set of adaptations. The scale of the experiment is limited, due to the cost of involving human testers. 
However, it is comparable to the ones of the main related approaches~\cite{behrang:apptestmigrator:ASE:2019,lin:craftdroid:ASE:2019}.


Another threat relates to the selection of testers. The four
testers have testing experience, but they are not the developers of the
apps. We mitigate this issue by letting 
the testers get familiarity with the apps before asking them to design the tests.

A final threat relates to the statistical significance of the results. Since the evolutionary algorithm is inherently stochastic, multiple runs may yield different results. Since the evaluation of the results involved human participants, who can only evaluate few tests, we ask them evaluate a single result of \tool.


\section{Related Work}
\label{related}


%
\smallskip
\noindent
\mbox{\textbf{GUI Test Generation}} 
Existing generators of GUI tests~\cite{
	Zein:studyandroid:JSS:2016,Kong:android:IEEETR:2019} have two major limitations: \begin{inparaenum}[(i)] \item lack of domain knowledge~\cite{mao:crowd:ASE:2017}, and thus they may generate either unrealistic or semantically meaningless GUI tests~\cite{bozkurt:realistic:sose:2011}, \item lack of automated oracles, and thus they are able to only detect crashes or exceptions~\cite{Moran:crashes:ICST:2016,Zhao:crash:ICSE:2019}\end{inparaenum}.
\nametool addresses these limitations by generating semantic GUI tests and oracles adapted from manually-written GUI tests of similar apps. 

Researchers have exploited usage data to
improve GUI test generation~\cite{mao:crowd:ASE:2017,linares:mining:MSR:2015,Lu:usagedata:mobilesoft:2016,Fard:existing:ASE:2014}.
For example, \textsc{Polariz}~\cite{mao:crowd:ASE:2017} and \textsc{MonkeyLab}~\cite{linares:mining:MSR:2015} generate Android tests using GUI interaction patterns extracted from app usage data. Differently, \nametool fully adapts existing tests across apps maintaining the same semantics of the donor test.

Similarly to \nametool, the GUI test generators \textsc{Augusto}~\cite{Mariani:Augusto:ICSE:2018} and \textsc{AppFlow}~\cite{Hu:appflow:FSE:2018} 
exploit commonalities among GUI applications.
However, they do not aim to adapt tests across applications nor leverage
existing GUI tests. Instead, they rely on a set of manually-crafted GUI interaction patterns. 
Moreover, \textsc{AppFlow} recognizes common widgets using a semi-automated machine learning approach. Conversely, \nametool matches GUI events without requiring human intervention.


\smallskip
\noindent
\mbox{\textbf{GUI Test Adaptation}} \textsc{CraftDroid}~\cite{lin:craftdroid:ASE:2019} and
\textsc{AppTestMigrator}~\cite{behrang:apptestmigrator:ASE:2019,Behrang:migration:ICSE:2018} are the first  attempts to adapt GUI tests across mobile apps.
Both approaches explore a statically computed GUI model of the \emph{recipient app} to ``greedly'' find a sequence of events that maximizes the semantic similarity with the events of the donor tests.
Similarly to \nametool, they extract event descriptors from the GUI and match them across applications using  word embedding~\cite{Mikolov:VectorSum:NIPS:2013}. 
However, \tool differs substantially from these two techniques.

\tool shares the overall objective of adapting tests across applications, but introduces 
substantial novelties.
Both \textsc{CraftDroid} and
\textsc{AppTestMigrator}  use a greedy algorithm, which resembles the one that \nametool uses to generate the initial population $\mathcal{P}_0$. 
\nametool uses an evolutionary algorithm to improve an initial set of greedy-matched tests.
As discussed in Section~\ref{empirical}, \nametool evolutionary approach largely improves over a greedy algorithm. 
Indeed, \textsc{AppTestMigrator} and \textsc{CraftDroid} explore \emph{a~single} test adaptation, and do not consider alternative sequences of random events that could yield to a better adaptation.
The \nametool evolutionary approach explores \emph{many possible}  test adaptations to find the sequence of events that yields the best adaptation. 
As such, \nametool can be used to improve the tests adapted with \textsc{AppTestMigrator} or \textsc{CraftDroid}.


\smallskip
\noindent
\textbf{GUI Test Repair} 
Test repair techniques fix tests that become invalid during software evolution~\cite{memon:adapt:2016:TSE,memon:automaticallyrepearing:tosem:2008,memon:regressiongui:jsme:2005,zhang:borken:issta:2013,Mirzaaghaei:TCA:ICST:2012,Li:atom:ICST:2017}. 
These techniques assume that most widgets remain unmodified between versions of the same app~\cite{Behrang:migra:ISSTA:2018,memon:adapt:2016:TSE}, and do not address the core challenge of semantically matching widgets across  apps.


\section{Conclusions}
\label{conclusions}

This paper presents \nametool an evolutionary technique to adapt test cases across mobile apps that share similar functionalities.
Our empirical evaluation indicates that \nametool can adapt useful and non-trivial GUI tests across semantically similar apps with very different GUIs.
This confirms that formulating the test adaptation problem with an evolutionary approach is a viable solution. 
An important future work is to reduce the computational cost of \nametool by implementing a distributed version of the tool that executes the evolutionary algorithm on the cloud.
Indeed, one of the key advantages of evolutionary algorithms is that are easily parallelizable.
Another interesting future work is to extend \nametool to adapt tests across different platforms, for instance, to adapt GUI tests from Mobile to Web applications.

\bibliographystyle{IEEEtran}
\bibliography{./bibliography/bibliography.bib,localbib.bib}

\end{document}